\documentclass[aip,reprint,pop]{revtex4-1}

\usepackage{graphicx}
\usepackage{subcaption}
\usepackage{amsmath}

\draft

\begin{document}

\title{Optimized stability of a modulated driver in a plasma wakefield accelerator} %Title of paper

\author{Roberto Martorelli}
\affiliation{Heinrich Heine Universit\"{a}t, 40225 D\"{u}sseldorf, Germany}

\author{Alexander Pukhov}
\affiliation{Heinrich Heine Universit\"{a}t, 40225 D\"{u}sseldorf, Germany}

\date{\today}

\begin{abstract}
We analyze the transverse stability for a configuration of multiple gaussian bunches subject to the self-generated plasma wakefield. Through a semi-analytical approach
we first study the equilibrium configuration for the modulated beam and then we investigate the evolution of the equilibrium configuration due to the emittance-driven
expansion of the beam front that results in a rigid backward shift. The rear-directed shift brings the modulated beam out of the equilibrium, with the possibility for
some of the bunch particles to be lost with a consequent deterioration of the driver. We look therefore for the proper position of the single bunches that maximize the
stability without severely affecting the accelerating field behind the driver. We then compare the results with 3D PIC simulations.
\end{abstract}


\maketitle

\section{Introduction}
\label{Sec.1}
Plasma wakefield acceleration, both laser driven\cite{Mangles, Faure, Geddes, Leemans} and particle driven \cite{Hogan, Ian}, has proven to be a possible alternative
to conventional accelerators for a wide range of applications, due to the intense electric fields that the plasma can sustain.\par
Among the possible configurations feasible for beam driven plasma wakefield, a renewed interest is devoted to the use of modulated drivers in view of the forthcoming
AWAKE experiment at CERN\cite{Awake}. The goal of the project is to prove the possibility to employ the $400$ GeV proton bunches produced at the Super Proton
Synchrotron(SPS) as drivers for the plasma wakefield.\par
The success of the experiment relies on the capability to produce short bunches whose length is approximately $L=\lambda_p/2$ (being $\lambda_p=2\pi c/\omega_p$ the
plasma wavelength and $\omega_p^2=4\pi ne^2/m$ the plasma frequency) from the much longer proton bunches provided at the SPS. Such achievement can be fulfilled through
the self-modulation instability\cite{Kumar}, in which the interaction of a long bunch with the wave itself leads to a modulation of the driver resulting at the final
stage in a train of bunches with the required length and periodicity approximately $\lambda_p$.\par
One of the serious issues of such a framework, and more in general for a configuration employing a pre-modulated beam, is the stability of the driver. A periodicity of
$\lambda_p$, although provides an intense accelerating field, does not guarantee as well a long lasting driver, which can be deteriorated by the transverse wake field.
Dealing with such a problem requires an appropriate analysis of the evolution of a modulated train of bunches interacting with a plasma.\par
In this paper we study the proper periodicity of the modulated driver in order to improve its stability without severely affecting the resulting accelerating field.
We show that by properly placing the single bunches it is possible to reduce the deterioration of the modulated configuration, providing a larger number of bunch
particles contributing to the wakefield for a longer time. This will end up eventually in a more intense accelerating field as compared to the case of a modulation of
$\lambda_p$, after few meters of propagation.\par
Such result suggests the necessity to rethink the proper shape of the modulated beam. While in the case of a pre-modulated driver it is enough to properly configure
the spacing between the bunches, for a self-modulated driver like in the AWAKE experiment, a different approach is necessary. A possible solution is the use of two
plasma cells\cite{Path}, the first intended to modulate the proton bunch through the self-modulation instability, the second to employ the modulated structure to
excite the field suitable for the acceleration of the witness bunch.\par
The paper is structured as follows: in section \ref{Sec.2} we trace the model used for description of the interaction of Gaussian bunches with a plasma, assuming a linear
response and the quasi-static approximation; in section \ref{Sec.3} we describe the evolution of the equilibrium configuration for the bunches emerging from the model,
comparing the results with the case of flat-top bunches; in section \ref{Sec.4} we look for the proper position of the bunches in order to increase their stability
without seriously affecting the longitudinal electric field; in section \ref{Sec.5} we present our conclusions.

\section{Analytical model}
\label{Sec.2}
The model describing the interaction of a particle beam with plasma relies on the work of Kenigs and Jones\cite{Kenigs}.\par
The authors consider an axi-symmetric bunch linearly interacting with an overdense plasma with immobile ions. Since the plasma is overdense the beam is regarded as an
external perturbation. The analysis is developed in the co-moving frame defined by the variables $\xi=\beta ct-z$ and $\tau=t$ with $\beta=v_b/c\simeq1$, $v_b$ being
the beam velocity and $c$ the speed of light. Lastly the quasi-static approximation is assumed, providing $\partial_\tau\simeq0$.\par
The two-dimensional transverse and longitudinal fields arising from the interaction are then:
\begin{align}
 W(r,\xi)&=(E_r-\beta B_\theta)=4\pi k_p\int_0^\infty\int_0^\xi\frac{\partial\rho(r',\xi')}{\partial r'}\nonumber\\
 &r'\operatorname{I_1}(k_pr_<)\operatorname{K_1}(k_pr_>)\sin[k_p(\xi-\xi')]d\xi'dr'\label{eq1}\\
 E_z(r,\xi)&=-4\pi k_p^2\int_0^\infty\int_0^\xi\rho(\xi',r')r'\nonumber\\
 &\operatorname{I_0}(k_pr_<)\operatorname{K_0}(k_pr_>)\cos[k_p(\xi-\xi')]d\xi'dr'\label{eq2}
\end{align}
where $k_p=\omega_p/v_b$ is the plasma wavenumber, $\rho(r,\xi)$ is the bunch charge density, $\operatorname{I_{1/0}}$ and $\operatorname{K_{1/0}}$ are the modified
Bessel functions and $r_{</>}=\operatorname{min}/\operatorname{max}(r,r')$.\par
We study the fields excited by a bunch with a flat-top profile in the transverse direction:
\begin{equation}
 \rho(r,\xi)=n_bq_b\left(\frac{r_0}{r_b(\xi)}\right)^2\operatorname{H}(r_b(\xi)-r)f(\xi)\label{eq3}
\end{equation}
with $n_b$ being the peak bunch density, $q_b$ the bunch charge, $r_0$ the initial bunch radius, $r_b(\xi)$ the radius of the beam-envelope, $\operatorname{H}$ the
Heaviside function and $f(\xi)$ the longitudinal bunch profile. The wakefields generated by such a distribution are therefore:
\begin{align}
 &W(r,\xi)=-4\pi k_pn_bq_br_0^2\nonumber\\
 &\begin{cases}
 \operatorname{I_1}(k_pr)\int_0^\xi f(\xi^\prime)\frac{\operatorname{K_1}(k_pr_b(\xi^\prime))}{r_b(\xi^\prime)}\sin[k_p(\xi-\xi^\prime)]d\xi^\prime\text{ for }r<r_b\\
 \operatorname{K_1}(k_pr)\int_0^\xi f(\xi^\prime)\frac{\operatorname{I_1}(k_pr_b(\xi^\prime))}{r_b(\xi^\prime)}\sin[k_p(\xi-\xi^\prime)]d\xi^\prime\text{ for }r>r_b.\label{eq4}
 \end{cases}\\
 &E_z(r,\xi)=-4\pi k_pn_bq_br_0^2\nonumber\\
 &\begin{cases}
 \int_0^\xi f(\xi^\prime)\frac{1-k_pr_b(\xi')\operatorname{I_0}(k_pr)\operatorname{K_1}(k_pr_b(\xi^\prime))}{k_pr_b^2(\xi^\prime)}\cos[k_p(\xi-\xi^\prime)]d\xi^\prime\text{ for }r<r_b\\
 \operatorname{K_0}(k_pr)\int_0^\xi f(\xi^\prime)\frac{\operatorname{I_1}(k_pr_b(\xi^\prime))}{r_b(\xi^\prime)}\cos[k_p(\xi-\xi^\prime)]d\xi^\prime\text{ for }r>r_b.\label{eq5}
 \end{cases}
\end{align}
In order to self-consistently include the dynamics of the beam, we couple the fields equation with the beam-envelope equation for the beam radius assuming that the
transverse motion of the beam can be described by only the motion of its boundary like in a water-bag model.\par
Denoting with $r_b=r_b(\xi,\tau)$ and with $r_b'=r_b(\xi',\tau)$, the resulting equation is:
\begin{align}
 &\frac{\partial^2r_b}{\partial\tau^2}=\frac{\epsilon^2c^2}{\gamma^2r_b^3}-\frac{4\pi k_pn_bq_b^2r_0^2}{m_b\gamma}\nonumber\\
 &\begin{cases}
 \operatorname{I_1}(k_pr_b)\int_0^\xi f(\xi^\prime)\frac{\operatorname{K_1}(k_pr_b^\prime)}{r_b^\prime}\sin[k_p(\xi-\xi^\prime)]d\xi^\prime\text{ for } r_b<r_b'\\
 \operatorname{K_1}(k_pr_b)\int_0^\xi f(\xi^\prime)\frac{\operatorname{I_1}(k_pr_b^\prime)}{r_b^\prime}\sin[k_p(\xi-\xi^\prime)]d\xi^\prime\text{ for } r_b>r_b'
 \end{cases}\label{eq6}
\end{align}
with $\epsilon$ being the normalized beam emittance, $\gamma=1/\sqrt{1-\beta^2}$ the beam relativistic Lorentz factor and $m_b$ the mass of the beam particles.\par
From Eq.\ref{eq6} results that the front of the beam is not subject to the wakefield and its dynamics is governed by the emittance-driven expansion according to the
equation:
\begin{equation}
 r_b(\xi=0,\tau)=r_{b0}(\tau)=r_0\sqrt{1+\frac{\epsilon^2c^2\tau^2}{r_0^4\gamma^2}}.\label{eq7}
\end{equation}
The purely diverging front of the bunch provides the absence of a global stable configuration for the transverse beam profile. Moreover, since the dynamics at every
point along the bunch depends on the upstream part, the evolution of the front of the bunch leads to a change of the whole equilibrium configuration for the beam while
propagating in the plasma.\par

\section{Evolution of the equilibrium configuration for gaussian bunches}
\label{Sec.3}
A previous work\cite{Martorelli} has analyzed the equilibrium configuration for a modulated bunch with longitudinal flat-top density profile. The authors have shown
that the dynamics of the front of the bunch leads to a backward drift of the whole equilibrium configuration for the modulated beam.\par
We perform the same analysis for the case of bunches with a longitudinal gaussian density profile. Since the defocusing force driven by the emittance does not depend on
the explicit shape of the bunch, we can expect a similar behavior as well for gaussian bunches.\par
First we consider the case of a modulated beam composed by identical equidistant bunches. The bunch densities are described by Eq.(\ref{eq3}) with longitudinal profiles:
\begin{equation}
 f(\xi)=\sum_{j=1}^Ne^{-\frac{(\xi-\xi_j)^2}{2\sigma_j^2}}=\sum_{j=1}^Ne^{-\frac{(\xi-\xi_0-j\Delta)^2}{2\sigma^2}}
\end{equation}
with $N$ being the number of bunches, $\xi_j$ the center, $\sigma_j$ the length and $\Delta$ the periodicity. The complete set of the parameters characterizing both the
plasma and the beam, with the exclusion of the beam length, is based on the baseline of the AWAKE project and is presented in Table \ref{tab1}.\par
Although this analysis refers explicitly to the AWAKE experiment, the mechanism that leads to the modulation can be general. Therefore the same results hold for both a
pre-modulated and a self-modulated beam.\par
\begin{table}
 \caption{Simulation parameters for plasma and bunch.\label{tab1}}
 \begin{tabular}{| c | r |}
 \hline
 Parameter & Value\\ \hline
 Plasma density $(n_p)$ & $7\times10^{14}$ cm$^{-3}$\\ \hline
 Bunch length $(\sigma)$ & $0.02$ cm\\ \hline
 Initial bunch radius $(r_0)$ & $0.02$ cm\\ \hline
 Periodicity $(\Delta)$ & $0.12$ cm\\ \hline
 Peak bunch density $(n_b)$ & $4\times10^{12}$ cm$^{-3}$\\ \hline
 Bunch relativistic Lorentz factor $(\gamma)$ & 400\\ \hline
 Normalized bunch emittance $(\epsilon)$ & $3.5$ mm mrad\\ \hline
 \end{tabular}
\end{table}
The equilibrium radius of the bunches is defined as the global minimum, at every position in the bunch, of the potential provided by the force in Eq.\ref{eq6}, with the
prescription that $r_b(\xi,\tau)\leq r_b(\xi=0,\tau)$.\par
\begin{figure}
\includegraphics[scale=.8]{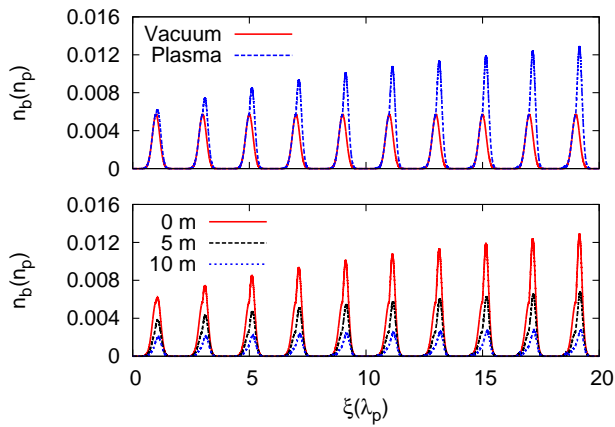}
\caption{Quasi-equilibrium configuration for a modulated gaussian bunch with periodicity $2\lambda_p$ in vacuum and in plasma (top) and for different propagation distances in
the plasma (bottom).\label{fig1}}
\end{figure}
As expected (Fig.\ref{fig1}), the interaction with the plasma leads to the focusing of the modulated beam, increasing the peak density of the bunches as compared to the
case of pure vacuum. The focusing force increases towards the tail of the configuration due to the interference among the transverse fields generated by the single
bunches.\par
\begin{figure}
\includegraphics[scale=.8]{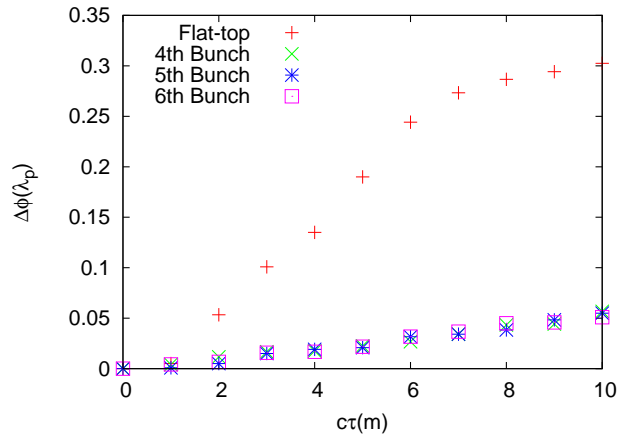}
\caption{Shift of the 4th, 5th and 6th bunch respect to its initial position varying the propagation distance in the plasma. The result is compared with the shift
experienced by flat-top bunches.\label{fig2}}
\end{figure}
\begin{figure}
\includegraphics[scale=.8]{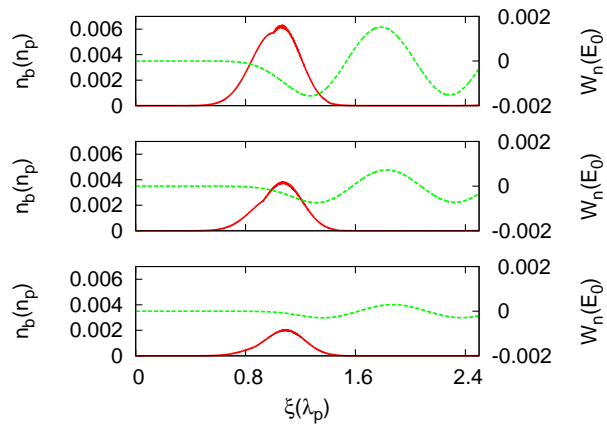}
\caption{Transverse field excited by a gaussian bunch for $0$, $5$ and $10$ m of propagation in the plasma.\label{fig3}}
\end{figure}

In order to avoid effects due to the overlapping of the single bunches, the periodicity has been set to $\Delta=2\lambda_p$. Studying the evolution of the equilibrium
configuration for gaussian bunches while propagating in the plasma, shows a behavior analogous to the case of flat-top bunches. The modulated beam tends to experience
a backward shift in its equilibrium configuration, with the displacement that is increasing with the propagation distance. Nevertheless, as shown in Fig.\ref{fig2}, the
shift is much smaller for gaussian bunches than that for flat-top ones.
Although the backward shift is quantitatively different for the two cases, the reason is the same. The expansion of the front of the bunch corresponds to a decreasing
bunch density and therefore a larger amount of bunch is required to obtain the same charge for different propagation distances. This process delays the onset of the
focusing force(Fig.\ref{fig3}). The inhomogeneity of the gaussian bunches tends anyway to a suppression of the fields, therefore the backward shift is less strong
respect to flat-top bunches.\par
The evolution of the equilibrium configuration implies serious consequences for the stability and the duration of the modulated beam.\par
A bunch particle initially in an equilibrium position, will find itself in an unstable one while propagating in the plasma as a consequence of the rearrangement of the
trapping potential caused by the expansion of the front.\par
The displacement experienced by the initial bunch radius with respect to the new equilibrium configuration results in a gain of transverse momenta. If the displacement
is large enough, the bunch particles can even escape the trapping potential.\par
The more driver particles are depleted, the lower is the wakefield amplitude, resulting in the lower efficiency of the process.

\section{Optimization of the beam configuration}
\label{Sec.4}
The framework depicted previously demands a further analysis on the optimal configuration for a modulated beam in order to reduce its deterioration as much as possible,
while propagating in the plasma.\par
A train of bunches with periodicity $\Delta=\lambda_p$ provides an intense accelerating field behind the driver, but the simple expansion of the front of the bunch
causes its degradation. The more particles the bunches lose while propagating in the plasma, the weaker is the final accelerating field.\par
We look therefore to the proper position of the single bunches in order to maximize the number of particles keeping trapped while propagating in the plasma.\par
The trapping is studied between the initial and final position in the plasma channel. Since the backward shift increases with the propagation distance, improving the
trapping respect to the final stage guarantees as well an improvement for the whole evolution.\par
To establish which sections of the bunches are unstable, we compare the initial configuration with that at the end of the propagation. The bunch particles are lost if
the difference between the two is large enough to provide the necessary transverse momentum to escape the trapping potential.
\begin{figure}
\includegraphics[scale=.8]{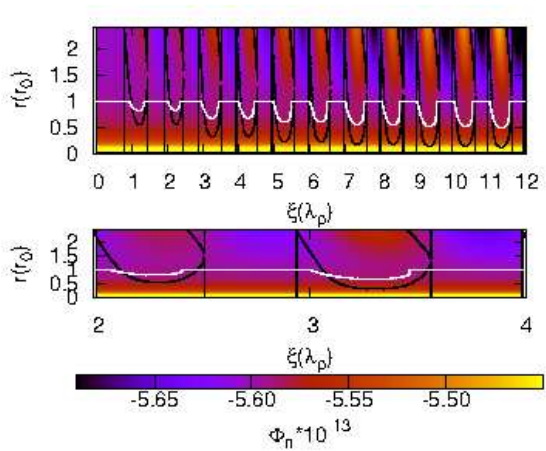}
\caption{Potential surface after $10$ m of propagation in the plasma. The white line corresponds to the equilibrium beam radius at $0$ m, while the black lines are
the boundaries for which the particle stay trapped after $10$ m of plasma.\label{fig4}}
\end{figure}
In Fig.\ref{fig4} we can see an example of the method applied: some sections of the initial equilibrium configuration are, at the end of the propagation distance, out
of the trapping region, meaning that those slices of the bunches are lost.\par
According to this criterion we look for the matching periodicity of the configuration that guarantees the highest number of particles trapped between the initial and
final step. The procedure is performed for one bunch at the time, fixing the configuration upstream to the already evaluated matching positions.\par
\begin{figure}
\includegraphics[scale=.8]{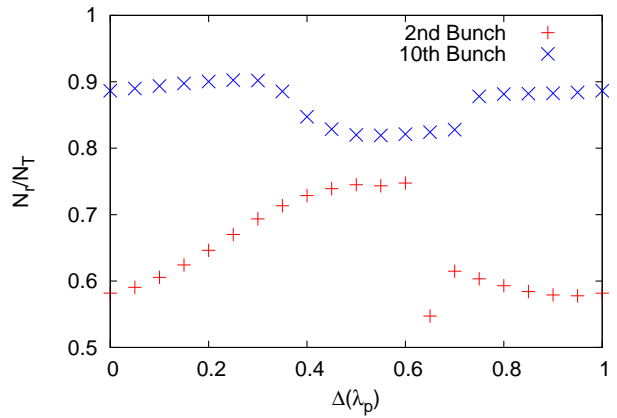}
\caption{Relative number of particle trapped after $10$ m of propagation varying the position of the last bunch of the configuration.\label{fig5}}
\end{figure}

In Fig.\ref{fig5} we can see the relative number of particle trapped changing the position of the last bunch of the configuration over a plasma wavelength.\par
As expected the maximum number of particle trapped is not achieved with a periodicity $\Delta=\lambda_p$. This is a consequence of several aspects: the maximum of the
Green function for the transverse field in Eq(\ref{eq2}) lies at $\Delta=\lambda_p/2$; the gaussian profile induces a nonlinear shift of the maximum; the backward drift
driven by the expansion of the front of the bunch causes an additional shift of the optimal position.\par
\begin{figure}
\includegraphics[scale=.8]{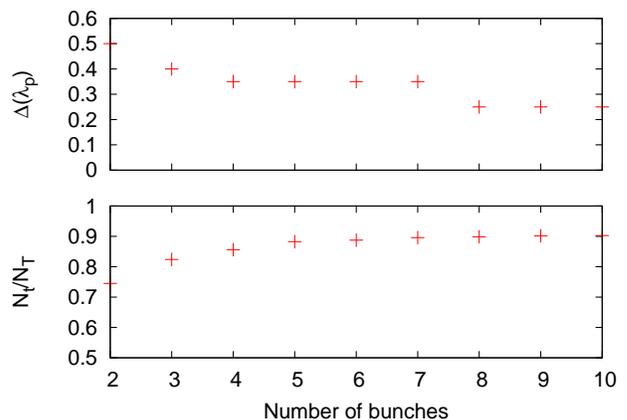}
\caption{Matching position of every bunch to improve the trapping (top) and relative number of particle trapped (bottom).\label{fig6}}
\end{figure}

Moreover, as appears from Fig.\ref{fig6}, the displacement is not constant for every bunch, meaning that it is not enough to rigidly move the configuration backward, but
an analysis for every bunch is required. On the other hand the stability of the configuration, defined as the relative number of particle trapped, increases with the
number of bunches involved. This is a consequence of the increasing total transverse field due to the superposition of the single ones generated by all the upstream
bunches.
Although this optimization provides an increasing final total charge of the beam, it does not necessarily implies as well an intense accelerating field behind the driver.
The matching position improving the total charge can coincide with that providing a destructive interference of the accelerating field generated by the single bunches.\par
This can be understood also by looking at Eq.(\ref{eq2}), since the Green function for the longitudinal and transverse fields are $\pi/2$ out of phase, meaning that the
maximum of the first will be close to the minimum of the second.\par
Therefore in order to obtain a configuration suitable for particle acceleration, it is necessary to improve the stability of the train of bunches without severely
affecting the resulting longitudinal field behind the driver.\par
We perform an analogous analysis as the previous one, taking into account this time also the average of the accelerating field behind the driver over the
propagation distance. The analysis is performed again for one bunch at the time, fixing the configuration upstream to the matching one.\par
\begin{figure}
\includegraphics[scale=.8]{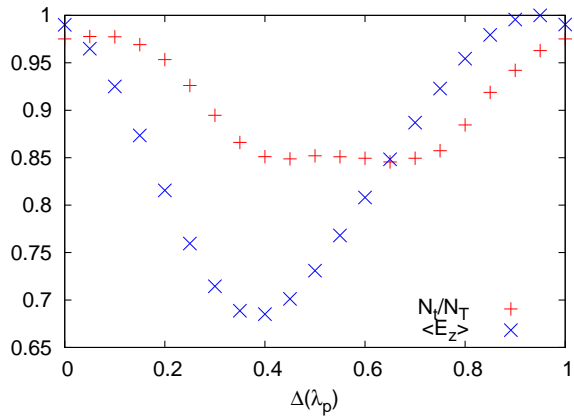}
\caption{Relative number of trapped particle with average accelerating field behind the driver for a configuration with ten bunches.\label{fig7}}
\end{figure}

Fig.\ref{fig7} displays clearly the behavior mentioned previously. The position that guarantees the maximum number of trapped particles does not correspond to that
improving also the longitudinal field behind the driver. In order to obtain the best of the two behaviors, the matching condition is set at the crossing between the
lines representing the relative number of trapped particle and the average accelerating field behind the driver.\par
\begin{figure}
\includegraphics[scale=.8]{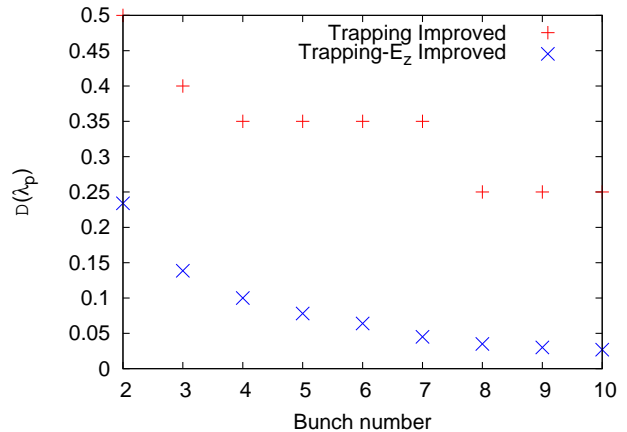}
\caption{Matching position of every bunch in order to improve both the stability of the driver and the final accelerating field. The result is compared with the previous one
in which only the stability was guaranteed.\label{fig8}}
\end{figure}
\begin{figure*}
\includegraphics[scale=.27]{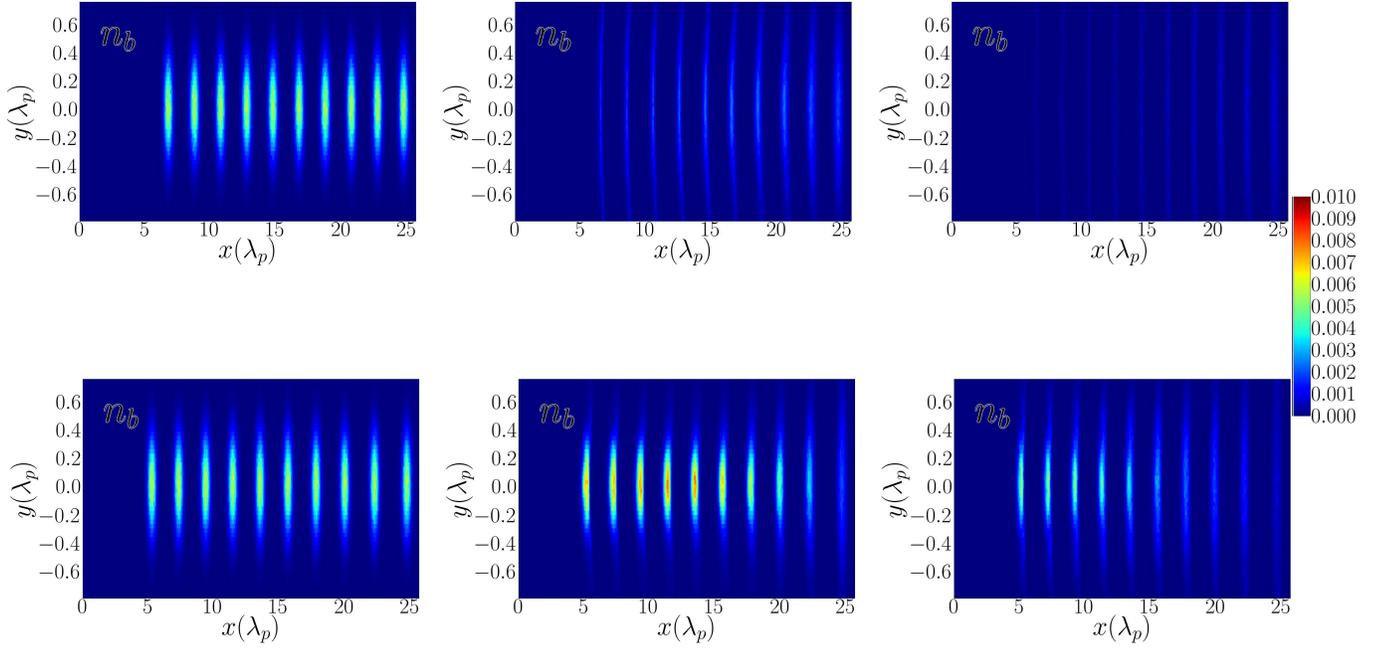}
\caption{Density of the modulated beam for a periodicity $\Delta=\lambda_p$ (upper line) and for the matching periodicity (lower line) at $0$, $5$ and $10$ m in the
plasma.\label{fig9}}
\end{figure*}

In order to confirm the results, we have performed a three dimensional PIC simulation using the quasi-static VLPL code\cite{Pukhov}. We compare both the bunch densities
and the longitudinal electric field for a modulation of $\lambda_p$ and the newly obtained modulation. We can see from Fig.\ref{fig9} that the optimized configuration
preserves the bunch densities, obtaining a final charge higher than that in the periodic case.\par
\begin{figure*}
\includegraphics[scale=.27]{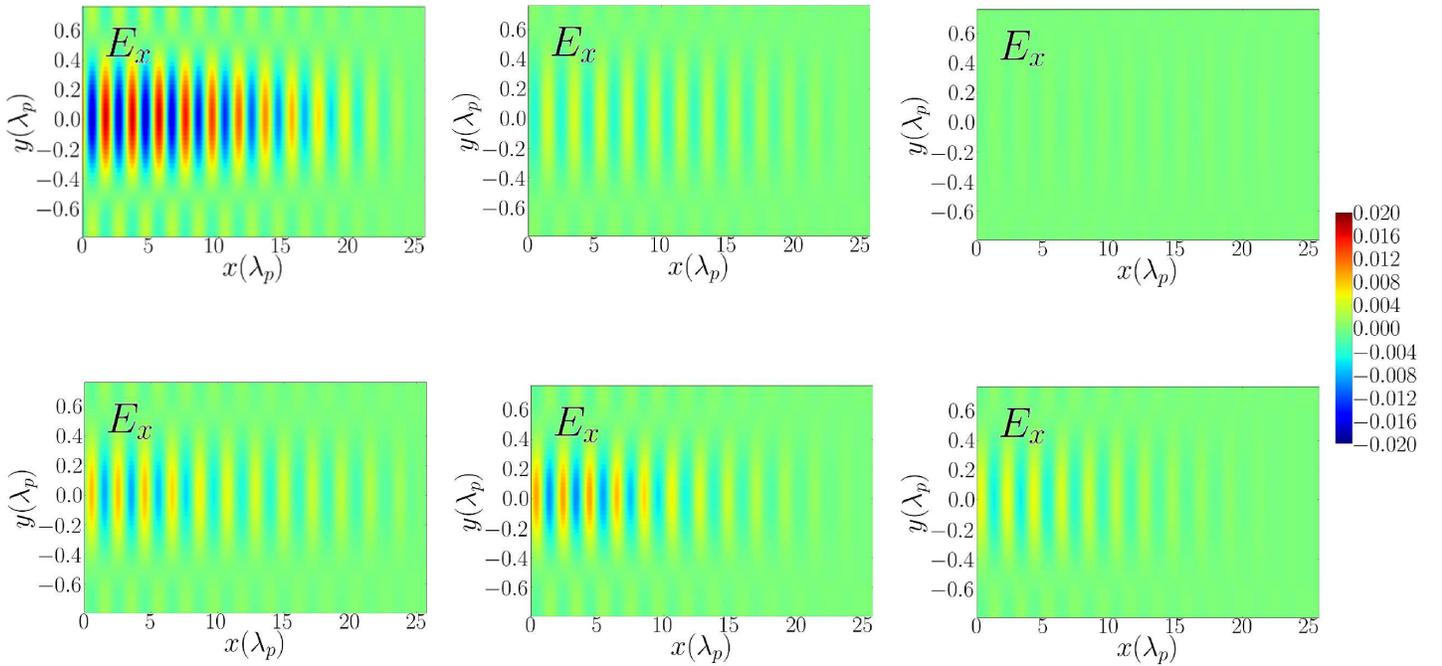}
\caption{Longitudinal electric field generated by the modulated beam for a periodicity $\Delta=\lambda_p$ (upper line) and for the matching periodicity (lower line) at $0$, $5$ and $10$ m in the
plasma.\label{fig10}}
\end{figure*}

\begin{figure}
\includegraphics[scale=.8]{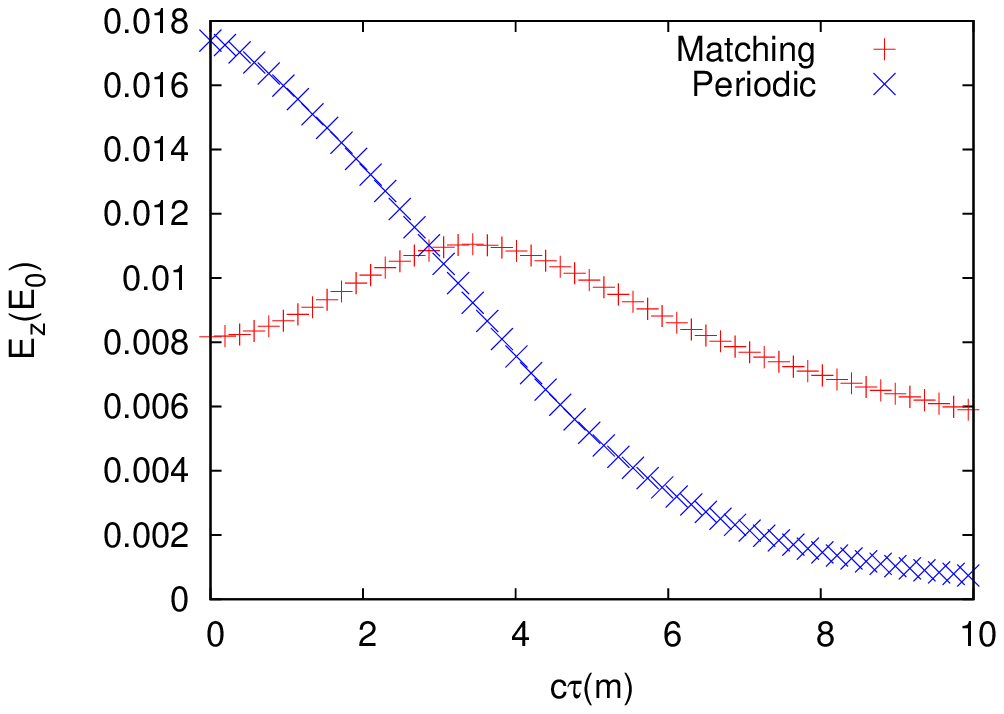}
\caption{Comparison of the accelerating field behind the driver for the periodic and optimized case as a function of the propagation distance.\label{fig11}}
\end{figure}

The longitudinal electric field as well provides the behavior expected by the improved configuration(Fig\ref{fig10}). The modulation of $\lambda_p$ guarantees a stronger
electric field initially. On the other hand, the optimized configuration provides an improved focusing field acting on the single bunches. This leads to an increasing
peak density and therefore an increasing accelerating field. After about $3$ meters of propagation in the plasma, the accelerating field behind the driver becomes
stronger in the new configuration(Fig.\ref{fig11}) and the improved stability shows guarantees a slower
decrease.\par

\section{Conclusion}
\label{Sec.5}
In this work we have analyzed the stability of a train of gaussian bunches interacting with a plasma, a configuration occurring in plasma wakefield acceleration
employing pre-modulated or self-modulated beams.\par
Through a semi-analytical model we have study the equilibrium configuration for a series of periodic gaussian bunches and the effects of the expansion of the front of
the bunch in its evolution. The result shows an analogy with the backward shift occurring for flat-top bunches. To excite focusing fields of the same intensity for
increasing propagation distances, it is required a longer section of the bunch, due to the decreasing density of the front. This provides a phase shift of the focusing
field and therefore of the entire equilibrium configuration.\par
We pointed out as this backward shift leads to an instability of the driver and to its deterioration while propagating in the plasma.\par
Through a numerical analysis we have found first the proper position of the bunches in order to improve the relative number of particle trapped during the propagation
showing that is not provided by a periodicity of $\lambda_p$. We finally extended the analysis checking as well the resulting accelerating field behind the driver generated
by the new configuration. The final result provides a configuration for a train of gaussian bunches with an improved stability as well with a strong accelerating field
behind the driver.\par
The validity of the analysis have been finally tested performing a 3D PIC simulation using the quasi-static VLPL code, providing the effects already observed in the
simplified model. The accelerating field arising from the modified train of bunches is initially less intense than that emerging from a modulation of $\lambda_p$, but
after about $3$ meters of propagation it becomes stronger, the difference between the two increasing further for the rest of the propagation distance.

\begin{acknowledgments}
This work was founded by DFG TR18 and EuCARD$^2$. The authors would like to thank Dr. J. P. Farmer for useful discussions on the numerical aspects developed in this
work.
\end{acknowledgments}

\bibliography{TransverseStability}

\end{document}